\documentclass{elsarticle}

\usepackage{graphicx}
\usepackage{amsmath}
\usepackage{amssymb}
\usepackage{algorithm}
\usepackage{algpseudocode}
\usepackage{color}
\usepackage{colortbl}

\newtheorem{thm}{Theorem}

\newdefinition{rmk}{Remark}
\newproof{pf}{Proof}
\newtheorem{definition}{Definition}

\newcommand{\argmin}{\textnormal{arg}\min}
\newcommand{\Succ}{{\mathcal S}}
\newcommand{\Out}{{\mathcal O}}
\newcommand{\R}{{\mathbb{R}}}
\newcommand{\N}{{\mathbb{N}}}

\begin{document}

\begin{frontmatter}

\title{Low-Complexity Quantized Switching Controllers\\ using Approximate Bisimulation\tnoteref{thanks}}
\tnotetext[thanks]{This work was supported by the Agence Nationale de la Recherche (VEDECY project - ANR 2009 SEGI 015 01) and by the pole MSTIC of Universit\'e Joseph Fourier (SYMBAD project).}

\author[ljk]{Antoine Girard}
\ead{Antoine.Girard@imag.fr}

\address[ljk]{Laboratoire Jean Kuntzmann, Universit\'e Joseph Fourier,\\
51 rue des Math\'ematiques, B.P. 53,
38041 Grenoble Cedex 9, France}


\begin{abstract} 
In this paper, we consider the problem of synthesizing low-complexity controllers for incrementally stable switched systems.
For that purpose, we establish a new approximation result for the computation of symbolic models that are approximately bisimilar to a given switched system.
The main advantage over existing results is that it allows us to design naturally quantized switching controllers for safety or reachability specifications; 
these can be pre-computed offline and therefore the online execution time is reduced. Then, we present a technique to reduce the memory needed to store 
the control law by borrowing ideas from algebraic decision diagrams for compact function representation and by exploiting the non-determinism of the synthesized controllers. We show the merits of our approach by applying it to a simple model of temperature regulation in a building.
\end{abstract}

\end{frontmatter}

\section{Introduction}
The use of discrete abstractions or symbolic models has become quite popular for hybrid systems design 
(see e.g.~\cite{raisch1998,moor1999,tabuada2006,KloetzerB06,reissig2009}).
In particular, several recent works have focused on the use of symbolic models related to the original system by approximate equivalence relationships
(approximate bisimulations~\cite{tazaki08,B-GirPolTab08}; or approximate alternating simulation or bisimulation 
relations~\cite{A-PolTab09,mazo2010}) which give more flexibility in the abstraction process by allowing the observed behaviors of the symbolic model and of the original system to be different provided they remain close. These approximate behavioral relationships have enabled the development of new abstraction-based controller
synthesis techniques~\cite{tabuada2009,Girard2012}.

In this paper, we go one step further by pursuing the goal of synthesizing controllers of lower complexity with shorter execution time and 
more efficient memory usage for their encoding. For that purpose, we establish a new approximation result for the computation of symbolic models that are approximately bisimilar to a given incrementally stable switched system. This result is the first main contribution of the paper, it differs from the original result presented by~\cite{B-GirPolTab08} mainly by the fact that the expression of the approximate bisimulation relation uses a quantized value of the state of the switched system rather than its full value in~\cite{B-GirPolTab08}.
This difference is fundamental for the synthesis of controllers with lower complexity. Indeed, the combination of this new result
with synthesis techniques for safety or reachability specifications presented in~\cite{Girard2012} yields quantized switching controllers that can be entirely pre-computed offline.
The online execution time is then greatly reduced in comparison to controllers obtained using the previous existing approximation result.
The second main contribution of the paper is to consider the problem of the representation of the control law with the goal of reducing the memory needed for its storage.
This is done by using ideas from algebraic decision diagrams (see e.g.~\cite{ADD}) for compact function representation.
Also, the non-determinism of the synthesized controllers can be exploited to further simplify the representation of the control law.
Finally, we apply our approach to the synthesis of controllers for a simple model of temperature regulation in a building.
The results on the synthesis of safety controllers appeared in preliminary form in the conference paper~\cite{girard2012a}, those on reachability controllers are new.

\section{Symbolic Models for Switched Systems}

In this section, we present an approach for the computation of symbolic models (i.e. discrete abstractions)
for a class of switched systems. This problem has been already considered by~\cite{B-GirPolTab08}.
In the following, we present a slightly different abstraction result that will allow us to synthesize controllers with lower complexity.

\subsection{Switched systems}

In this paper, we consider a class of switched systems of the form:
$$
\Sigma:\; \dot{\bf x}(t)=f_{{\bf p}(t)}({\bf x}(t)),\; {\bf x}(t) \in \R^{n}, \; {\bf p}(t) \in P
$$
where $P$ is a finite set of modes. 
The switching signals ${\bf p}:\R^+ \rightarrow P$ are assumed to be
piecewise constant functions, continuous from the right and with a finite number of discontinuities on every bounded interval. 
We use $\mathbf{x}(t,x,\mathbf{p})$ to denote the point
reached at time $t\in \R_0^+$ from the initial condition $x$ under
the switching signal $\mathbf{p}$. 
We will assume that the switched system $\Sigma$ is 
incrementally globally uniformly asymptotically stable~\cite{B-GirPolTab08}:
\begin{definition} The switched system $\Sigma$ is said to be
incrementally globally uniformly asymptotically stable ($\delta$-GUAS) if
there exists a ${KL}$ function\footnote{A continuous function $\gamma :\R^+_0 \rightarrow \R^+_0 $ is said
to belong to class $K_\infty$ if it is strictly increasing,
$\gamma(0)=0$ and
$\gamma(r)\rightarrow \infty$ when $r \rightarrow \infty$. A
continuous function $\beta:\R^+_0 \times \R^+_0 \rightarrow
\R^+_0$ is said to belong to class ${KL}$ if for all fixed
$s$, the map $r\mapsto \beta(r,s)$ belongs to class $
K_\infty$ and for all fixed $r$, the map $s \mapsto \beta(r,s)$ is
strictly decreasing and $\beta(r,s)\rightarrow 0$ when
$s\rightarrow \infty$.} $\beta$ such that for all
$t\in \R_0^+$, for all $x,y \in \R^n$, for all switching signals $\mathbf{p} \in \mathcal P$, the following condition is
satisfied:
\begin{equation}
\label{eq:stab2}
 \|\mathbf{x}(t,x,\mathbf p)-\mathbf{x}(t,y,\mathbf p)\| \le
\beta(\|x-y\|,t).
\end{equation}
\end{definition}

Intuitively, a switched system is $\delta$-GUAS if the distance between any two trajectories associated with the same switching signal {\bf p}, but with different initial states, converges asymptotically to $0$.
Incremental stability of a switched system can be characterized using Lyapunov functions~\cite{B-GirPolTab08}:
\begin{definition}
\label{def:lyap}
A smooth function $\mathcal V:\R^n\times \R^n \rightarrow
\R^+_0$ is a common $\delta$-GUAS Lyapunov function
for $\Sigma$ if there exist $ K_\infty$ functions
$\underline{\alpha}$, $\overline{\alpha}$ and a real number $\kappa>0$ such that for all $x,y\in \R^n$, for all $p\in P$:
$$
\underline{\alpha}(\|x-y\|) \le \mathcal V(x,y) \le
\overline{\alpha}(\|x-y\|);\\
$$
$$
\frac{\partial \mathcal V}{\partial x}(x,y)\cdot f_p(x)+ \frac{\partial
\mathcal V}{\partial y}(x,y)\cdot f_p(y) \le -\kappa \mathcal V(x,y)
.
$$
\end{definition}

It has been shown in~\cite{B-GirPolTab08} that the existence of a common $\delta$-GUAS Lyapunov function ensures that the switched system $\Sigma$ is $\delta$-GUAS.


We now introduce the class of labeled transition systems which will serve as a common  modeling framework for
switched systems and symbolic models.

\begin{definition}
A {\it transition system} 
$T=(X,U,\Succ,Y,\Out)$
consists of: 
\begin{itemize}
\item a set of states $X$; 
\item a set of inputs $U$;
\item a (set-valued) transition map $\Succ:X\times U \rightarrow 2^X$;
\item a set of outputs $Y$;
\item and an output map $\Out:X\rightarrow Y$.
\end{itemize}
\end{definition}

$T$ is \textit{metric} if the
set of outputs $Y$ is equipped with a metric ${d}$.
If the set of states $X$ and inputs $U$ are finite or countable, $T$ is said {\it symbolic} or {\it discrete}.

An input $u\in U$ belongs to the set of {\it enabled inputs} at state $x$, denoted $\textrm{Enab}(x)$, if 
$\Succ(x,u) \ne \emptyset$.
If $\textrm{Enab}(x)\ne \emptyset$, then the state $x$ is said to be {\em non-blocking}, otherwise it said to be {\em blocking}. 
The system is said to be non-blocking if all states are non-blocking.
If for all $x\in X$ and for all $u\in \textrm{Enab}(x)$, $\Succ(x,u)$ has $1$ element then the transition system is said to be {\em deterministic}.

A {\it state trajectory} of $T$ is a finite or infinite sequence of states and inputs,
$\{(x^i,u^i)|\; i=0,\dots,N\}$ (we can have $N=+\infty$) where $x^{i+1}\in \Succ(x^i,u^i)$ for all 
$i=0,\dots,N-1$.
The associated {\it output trajectory} is the sequence of outputs $\{y^i|\; i=0,\dots,N\}$
where $y^i=\Out(x^i)$ for all $i=0,\dots,N$.

Given a switched system $\Sigma$ and a parameter $\tau >0$, we define a transition system $T_\tau(\Sigma)$ that describes trajectories of $\Sigma$ of duration $\tau$. This can be seen as a time sampling process,
which is natural when the switching in $\Sigma$ is to be determined by a periodic controller of period $\tau$.
Formally, $T_\tau(\Sigma)=(X_1,U,\Succ_1,Y,\Out_1)$
where the set of states is $X_1=\R^n$; the set of inputs is the set of modes $U=P$;
the deterministic transition map is given by $x_1'=\Succ_1(x_1,p)$ if and only if 
$$
x_1'=\mathbf{x}(\tau), \text{ where } \dot{\mathbf{x}}(t)=f_p(\mathbf{x}(t)),\; \mathbf{x}(0)=x_1,\; t\in[0,\tau];
$$
the set of outputs is $Y=\R^n$; and
the observation map $\Out_1$ is the identity map over $\R^n$.
$T_{\tau}(\Sigma)$ is non-blocking, deterministic and metric when the set of
observations $Y=\R^n$ is equipped with the Euclidean norm.

\subsection{Symbolic models}

In the following, we present a method to compute discrete abstractions for $T_\tau(\Sigma)$.
For that purpose, we consider approximate equivalence relationships for labeled transition systems 
defined by approximate bisimulation relations introduced in~\cite{A-GirPap07}.

\begin{definition} \label{Def-App-Bisim}
Let $T_i=(X_i,U,\Succ_i,Y,\Out_i)$, $i=1,2$, 
be metric labeled transition systems with the
same sets of inputs $U$ and outputs $Y$ equipped with the 
metric $d$.
Let $\varepsilon \ge 0$, a relation ${\mathcal R}_\varepsilon \subseteq X_1\times X_2$  is called an
$\varepsilon$-{\it approximate bisimulation relation} between $T_1$ and $T_2$, if for
all $(x_1,x_2)\in {\mathcal R}_\varepsilon$:
\begin{enumerate}
\item $d \left(\Out_1(x_1),\Out_2(x_2)\right) \le \varepsilon$, 
\item $\forall u\in \textrm{Enab}_1(x_1)$, $\forall x_1' \in \Succ_1(x_1,u)$, $\exists x_2' \in \Succ_2(x_2,u)$ 
such that $(x'_1,x'_2)\in {\mathcal R}_\varepsilon$.
\item $\forall u\in \textrm{Enab}_2(x_2)$, $\forall x_2' \in \Succ_2(x_2,u)$, $\exists x_1' \in \Succ_1(x_1,u)$ 
such that $(x'_1,x'_2)\in {\mathcal R}_\varepsilon$.
\end{enumerate}
$T_1$ and $T_2$ are approximately bisimilar with precision $\varepsilon$ (denoted $T_1\sim_\varepsilon
T_2$), if there exists ${\mathcal R}_\varepsilon$, an
$\varepsilon$-approximate bisimulation relation between $T_1$ and $T_2$, such
that for all $x_1\in X_1$, there exists $x_2\in X_2$ such that
$(x_1,x_2)\in {\mathcal R}_\varepsilon$, and 
conversely.
\end{definition}


We briefly describe an approach similar to that presented
in~\cite{B-GirPolTab08} for computing approximately bisimilar discrete abstractions of $T_\tau(\Sigma)$ (i.e. a discrete labeled transition system that is approximately bisimilar to $T_\tau(\Sigma)$).
We start by approximating the set of states $X_1=\R^n$ by a lattice:
$$
[\R^n]_{\eta}=\left\{q\in \R^n \,\,\left| \; q_{i}
=k_{i}\frac{2\eta}{\sqrt{n}},\; k_{i}\in\mathbb{Z},\;
i=1,\dots,n\right.\right\},
$$
where $q_i$ is the $i$-th coordinate of $q$ and $\eta>0$ is a state space discretization parameter. 
We associate a quantizer $Q_\eta:\R^n\rightarrow [\R^n]_{\eta}$ defined as follows $q=Q_\eta(x)$ if and only if
$$
\forall i=1,\dots,n,\;  q_i-\textstyle{\frac{\eta}{\sqrt{n}}} \le x_i < q_i+\textstyle{\frac{\eta}{\sqrt{n}}}.
$$
It is easy to check that for all $x\in \R^n$, 
$\|Q_\eta(x)-x\| \le \eta$. Given a subset $X\subseteq \R^n$ we denote $Q_\eta(X) = \{Q_\eta(x) | x\in X\}$. 

We can then define the abstraction of $T_\tau(\Sigma)$ as the transition system
$
T_{\tau,\eta}(\Sigma)=(X_2,U,\Succ_2,Y,\Out_2)
$,
where the set of states is $X_2=[\R^n]_{\eta}$; the set of labels remains the same $U=P$;
the transition relation is essentially obtained by quantizing the transition relation of $T_\tau(\Sigma)$:
$$
\forall x_2\in [\R^n]_\eta,\; \forall p \in P,\; \Succ_2(x_2,p)=Q_\eta(\Succ_1(x_2,p));
$$
the set of outputs remains the same $Y=\R^n$; and the observation map $\Out_2$ is given by $\Out_2(q)=q$.
Note that the transition system $T_{\tau,\eta}(\Sigma)$ is
discrete since its sets of states and actions are respectively countable and finite.
Moreover, it is non-blocking, deterministic and metric when the set of
observations $Y=\R^n$ is equipped with the Euclidean norm.

The approximate bisimilarity of  $T_\tau(\Sigma)$  and $T_{\tau,\eta}(\Sigma)$ is related
to the incremental stability of switched system $\Sigma$. In the following, we shall assume
that there exists a common $\delta$-GUAS Lyapunov function $\mathcal V$ for $\Sigma$.
We need  to make the supplementary assumption on the $\delta$-GUAS Lyapunov function that there exists a ${K}_\infty$ function $\gamma$ such that for all $x_1,x_2,y_1,y_2 \in \R^n$
\begin{equation}
\label{eq:assum}
 |\mathcal V(x_1,x_2)-\mathcal V(y_1,y_2)| \le \gamma(\|x_1-y_1\|+\|x_2-y_2\|).
\end{equation}
We can show that this assumption is not restrictive provided $\mathcal V$ is smooth and we are interested in the dynamics of $\Sigma$ on a compact subset of $\R^n$, which is often the case in practice. 

We are now able to present a new approximation result for determining an approximate bisimulation relation between $T_\tau(\Sigma)$  and $T_{\tau,\eta}(\Sigma)$:

\begin{thm}\label{th:symb} Consider a switched system $\Sigma$, time and state space sampling parameters $\tau, \eta >0$  and a desired
precision $\varepsilon >0$. If there exists a common $\delta$-GUAS Lyapunov function $\mathcal V$ for $\Sigma$ such that equation (\ref{eq:assum}) holds and 
\begin{equation}
\label{eq:simcond}
\varepsilon \ge \eta + \underline{\alpha}^{-1}\left(\frac{\gamma(2\eta)+\gamma(\eta)e^{-\kappa \tau}}{1-e^{-\kappa \tau}}  \right)
\end{equation}
then 
$$
\mathcal R_\varepsilon=\left\{(x_1,x_2)\in X_1\times X_2 |\; \mathcal V(Q_\eta( x_1),x_2) \le \underline{\alpha}(\varepsilon-\eta)   \right\}
$$
is an $\varepsilon$-approximate bisimulation relation between $T_{\tau}(\Sigma)$ and
$T_{\tau,\eta}(\Sigma)$. Moreover,  $T_{\tau}(\Sigma)\sim_{\varepsilon} T_{\tau,\eta}(\Sigma)$.
\end{thm}

\begin{pf}
Let $(x_1,x_2)\in \mathcal R_\varepsilon$, then 
\begin{eqnarray*}
\|x_1-x_2\| & \le & \|Q_\eta(x_1)-x_2\| + \eta \\
& \le &
\underline{\alpha}^{-1}\left(\mathcal V(Q_\eta(x_1),x_2)\right)+\eta \\
&\le &  \underline{\alpha}^{-1}\left(\underline{\alpha}(\varepsilon-\eta)\right)+\eta = \varepsilon.
\end{eqnarray*}
Thus, the first condition of Definition~\ref{Def-App-Bisim} holds. Let us remark that 
$\textrm{Enab}_1(x_1)=\textrm{Enab}_2(x_2)=P$ and since $T_\tau(\Sigma)$ and $T_{\tau,\eta}(\Sigma)$ are deterministic, the second and third conditions  of Definition~\ref{Def-App-Bisim} are equivalent. Then, let $p\in P$, let $x_1'=\Succ_1(x_1,p)$ and $x_2'=\Succ_2(x_2,p)$
then using the properties of $\delta$-GUAS Lyapunov function $\mathcal V$ we obtain
\begin{eqnarray*}
\mathcal V(Q_\eta(x_1'),x_2') & = & \mathcal V(Q_\eta(\Succ_1(x_1,p)),Q_\eta(\Succ_1(x_2,p))) \\
& \le & \mathcal V(\Succ_1(x_1,p),\Succ_1(x_2,p)) +  \gamma(2\eta) \\
& \le & e^{-\kappa \tau} \mathcal V(x_1,x_2) + \gamma(2\eta) \\
& \le & e^{-\kappa \tau} \left( \mathcal V(Q_\eta(x_1),x_2) + \gamma(\eta)\right) + \gamma(2\eta) \\
& \le &  e^{-\kappa \tau} \underline{\alpha}(\varepsilon-\eta) + \gamma(2\eta)+\gamma(\eta)e^{-\kappa \tau} \\
& \le & \underline{\alpha}(\varepsilon-\eta)
\end{eqnarray*}
by equation (\ref{eq:simcond}). It follows that $(x_1',x_2') \in \mathcal R_\varepsilon$ which is consequently an $\varepsilon$-approximate bisimulation relation between $T_{\tau}(\Sigma)$ and
$T_{\tau,\eta}(\Sigma)$. Now, let $x_1 \in \R^n$ and let $x_2\in [\R^n]_\eta$ given by $x_2 = Q_\eta(x_1)$. Then,
$\mathcal V(Q_\eta(x_1),x_2)=0$ and $(x_1,x_2) \in \mathcal R_\varepsilon$.
Conversely, let $x_2\in [\R^n]_\eta$ and let $x_1 \in \R^n$ given by $x_1=x_2$, let us remark that $Q_\eta(x_1)=x_2$ then
$\mathcal V(Q_\eta(x_1),x_2)=0$ and $(x_1,x_2) \in \mathcal R_\varepsilon$. Hence, it follows that
$T_{\tau}(\Sigma)\sim_{\varepsilon} T_{\tau,\eta}(\Sigma)$. $\blacksquare$
\end{pf}

We would like to point out that for given $\tau>0$ and $\varepsilon>0$, it is always possible to find $\eta >0$ such that
equation (\ref{eq:simcond}) holds. Hence, it is possible for any time sampling parameter $\tau>0$ to compute symbolic models for switched systems of arbitrary precision $\varepsilon>0$ by choosing a sufficiently small state space sampling parameter $\eta>0$.

We would like to emphasize the differences between Theorem~\ref{th:symb} and the original approximation result presented in~\cite{B-GirPolTab08}. The computation of the abstractions are essentially the same.
The main difference lies in the expression of the approximate bisimulation relation:
$(x_1,x_2) \in \mathcal R_\varepsilon$ if and only if $\mathcal V(x_1,x_2) \le \underline{\alpha}(\varepsilon)$ in~\cite{B-GirPolTab08}, instead of $\mathcal V(Q_\eta( x_1),x_2) \le \underline{\alpha}(\varepsilon-\eta)$
in Theorem~\ref{th:symb}.
We will see in the next section that this difference is fundamental as it will allow us to synthesize quantized controllers. It should also be noted that  the relations to be satisfied by the abstraction parameters, $\tau$, $\eta$ and $\varepsilon$ are different: for identical precision and time sampling parameters Theorem~\ref{th:symb} generally requires a finer state sampling parameter than the results presented in~\cite{B-GirPolTab08}.

\begin{rmk} When the switched system does not admit a common $\delta$-GUAS function, an approximation result was established 
in~\cite{B-GirPolTab08}, based on the use of multiple Lyapunov functions and under a minimum dwell-time assumption. A result similar to Theorem~\ref{th:symb} can also be established in that case.
\end{rmk}

In the remainder of the paper, we consider a switched system $\Sigma$ with time and state space sampling parameters $\tau$ and $\eta$.
We shall work with the labeled transition systems $T_\tau(\Sigma)$ and $T_{\tau,\eta}(\Sigma)$ and we shall assume that the assumptions of Theorem~\ref{th:symb} hold. 
We will denote for $x\in \R^n$, ${\mathcal R}_\varepsilon (x) =\{q \in [\R^n]_{\eta} |\; (x,q)\in {\mathcal R}_\varepsilon \}$.
We will also use the relation 
$$
\overline{\mathcal R}_\varepsilon=\left\{(q,q')\in [\R^n]_{\eta}\times [\R^n]_{\eta} |\; \mathcal V(q,q') \le \underline{\alpha}(\varepsilon-\eta)   \right\}
$$
and we denote for $q\in [\R^n]_{\eta}$, $\overline{\mathcal R}_\varepsilon (q) =\{q'\in [\R^n]_{\eta} |\; (q,q')\in \overline{\mathcal R}_\varepsilon \}$.
Let us remark that  for all $x\in \R^n$, ${\mathcal R}_\varepsilon (x) = \overline{\mathcal R}_\varepsilon(Q_\eta( x))$.

\section{Synthesis of Quantized Switching Controllers}

In this section, we present an approach for synthesizing quantized switching controllers for safety or reachability specifications.
It is based on the use of Theorem~\ref{th:symb} combined with controller synthesis techniques presented in~\cite{Girard2012}.
We start by defining the notion of controller for labeled transition systems:

\begin{definition} A controller for transition system $T=(X,U,\Succ,Y,\Out)$ is a set-valued map $\mathcal C: X\rightarrow 2^U$ such that $\mathcal C(x)\subseteq \textrm{Enab}(x)$, for all $x\in X$. 
The domain of $\mathcal C$ is the set $\textrm{dom}(\mathcal C)=\left\{x\in X|\; \mathcal C(x)\ne \emptyset\right\}$.
The dynamics of the controlled system is described by the transition system $T/\mathcal{C}=(X,U,\Succ_{\mathcal C},Y,\mathcal O)$ where the transition map is given by $x'\in \Succ_{\mathcal C}(x,u)$ if and only if $u\in \mathcal C(x)$ and
$x'\in \Succ(x,u)$.
\end{definition}

We would like to emphasize the fact that the controllers are set-valued maps, at a given state $x$ it enables a set of
admissible inputs $\mathcal C(x) \subseteq U$. A controller essentially executes as follows.
The state $x$ of $T$ is measured, an input $u\in \mathcal C(x)$ is selected and actuated. Then, the system takes a transition $x'\in \Succ(x,u)$. 
The blocking states of $T/{\mathcal C}$ are the elements of $X\setminus \textrm{dom}(\mathcal C)$.
Given a subset $X'\subseteq X$, we denote ${\mathcal C}(X')=\bigcup_{x\in X'} {\mathcal C}(x)$.

\subsection{Safety controllers}


Let $Y_S \subseteq Y$ be a set of outputs associated with safe states. 
We consider the safety synthesis problem that consists in determining a controller that keeps the output of the system inside the specified safe set $Y_S$. 

\begin{definition}\label{def:safe} Let $Y_S\subseteq Y$ be a set of {\textit safe} outputs. A controller $\mathcal C$ is a safety controller for $T=(X,U,\Succ,Y,\Out)$ and specification $Y_S$ if for all $x\in\textrm{dom}(\mathcal C)$:
\begin{enumerate}
\item $\Out(x) \in Y_S$ (safety);
\item $\forall u\in \mathcal C(x)$, $\Succ(x,u)\subseteq \textrm{dom}(\mathcal C)$ (deadend freedom).
\end{enumerate}
\end{definition}

It is easy to verify from the previous definition that for any initial state $x^0\in \textrm{dom}(\mathcal C)$, the controlled system $T/{\mathcal C}$ will never reach a blocking state (because of the deadend freedom condition) and its outputs will remain
in the safe set $Y_S$ forever (because of the safety condition). 

We now consider the problem of synthesizing a safety controller for $T_\tau(\Sigma)$ describing the sampled dynamics of the switched system $\Sigma$.
Let us consider a safety specification given by a compact set $Y_S \subseteq \R^n$.
We shall use a method developed in~\cite{Girard2012} for synthesizing safety controllers for labeled transition systems
using approximately bisimilar abstractions. Let us define the $\varepsilon$-contraction of $Y_S$ as 
$$
\textrm{Cont}_\varepsilon (Y_S)=
\left\{y \in Y_S |\; \forall y' \in \R^n, \|y-y'\|\le \varepsilon \Rightarrow y'\in Y_S \right\}.
$$

\begin{thm}\label{th:quantized} 
Let $\mathcal K_\varepsilon : [\R^n]_\eta \rightarrow 2^P$ be a safety controller for the symbolic model $T_{\tau,\eta}(\Sigma)$ and specification $\textrm{Cont}_\varepsilon (Y_S)$. Let $\mathcal K: [\R^n]_\eta \rightarrow 2^P$ be given for $q\in [\R^n]_\eta$ by 
\begin{equation}
\label{eq:ref}
\mathcal K(q) =  
\mathcal K_\varepsilon \left(\overline{\mathcal R}_\varepsilon (q)\right).
\end{equation}
Then, the map $\mathcal C:\R^n \rightarrow 2^P$ given by 
$\mathcal C =\mathcal K \circ Q_\eta$
is a safety controller for $T_\tau(\Sigma)$ and specification $Y_S$.
\end{thm}

\begin{pf}
By Theorem~1 in~\cite{Girard2012}, we have that $\mathcal C : \R^n \rightarrow 2^P$ given by
$\mathcal C(x)=  \mathcal K_\varepsilon ( \mathcal R_\varepsilon(x))
$
is a safety controller for $T_\tau(\Sigma)$ and specification $Y_S$. Then, using the 
fact that $\mathcal R_\varepsilon(x)= \overline{\mathcal R}_\varepsilon(Q_\eta(x))$ we obtain $\mathcal C=\mathcal K \circ Q_\eta$.
$\blacksquare$
\end{pf}

 It is to be noted that since $Y_S$ is compact, the set of states of the symbolic model
$T_{\tau,\eta}(\Sigma)$ with associated outputs in $\textrm{Cont}_\varepsilon (Y_S)$ is finite. As a consequence, the synthesis of the safety controller $\mathcal K_\varepsilon$ can be done by a simple fixed-point algorithm which is guaranteed to terminate in a finite number of steps (see e.g.~\cite{tabuada2009} for details). 


%

Let us remark that the only non-trivial values of $\mathcal C(x)$ are for $x\in Y_S$ since from a state $x\notin Y_S$, the safety specification cannot be met and therefore $\mathcal C(x)=\emptyset$. Hence, it is only necessary to compute $\mathcal K$ on $Q_\eta(Y_S)$ which is finite since
 $Y_S$ is a compact subset of $\R^n$.
Hence, it is possible to entirely pre-compute offline the discrete map $\mathcal K$. Then, for a state $x\in \R^n$ the computation
of the inputs enabled by $\mathcal C$ only requires quantizing the state $x$ and evaluating $\mathcal K(Q_\eta(x))$.
Thus, Theorem~\ref{th:quantized} gives an effective way to compute a {quantized} safety controller
for $T_\tau(\Sigma)$. Moreover, as shown in~\cite{Girard2012}, it is possible to give guarantees on the distance between the synthesized controller $\mathcal C$ and the most permissive  controller for the safety specification $Y_S$.

Let us now discuss the complexity of the synthesized controller\footnote{In the following, the notations $O(.)$ must be understood as asymptotic upper-bound estimates when $\eta$ approaches $0$.}.
The online execution time of the controller defined in Theorem~\ref{th:quantized} is in $O(n)$ (cost of a quantization)
and does not depend on the state space sampling parameter $\eta$. 
However, the memory space needed to store naively the control law (that is the map $\mathcal K$) is proportional to the number of states 
in $Q_\eta(Y_S)$, that is $O(\eta^{-n})$ which can be quite large in practice.
In comparison, using the approximate bisimulation relation given in~\cite{B-GirPolTab08} and Theorem~1 in (\cite{Girard2012}), the synthesized controller
would have been given by 
$$
\mathcal C(x)=\bigcup_{q'\in [\R^n]_\eta,\; \mathcal V(x,q')\le \underline{\alpha}(\varepsilon)}  
\mathcal K_\varepsilon (q').
$$
It is to be noted that the continuous state $x$ is not quantized and therefore the union cannot be computed offline for all possible values of $x$ as previously but has to be computed online.
In practice, the number of elements $q'\in [\R^n]_\eta$ such that 
$\mathcal V(x,q')\le \underline{\alpha}(\varepsilon)$ is in $O(({\varepsilon}/{\eta})^{n})$ which can be quite large.
Also the memory space needed for the storage of the map $\mathcal K_\varepsilon$
is also in $O(\eta^{-n})$.
Hence, we can see that our new approximation result allows us to synthesize controllers with smaller execution time and comparable memory usage.

\subsection{Reachability controllers}

Let $Y_S\subseteq Y$ be a set of outputs associated with safe states, let $Y_T\subseteq Y_S$ be a set of outputs associated with target states. 
We consider the reachability synthesis problem that consists in determining a controller steering the output of the system to $Y_T$ while keeping the output in $Y_S$ along the way. For simplicity, we assume that the labeled transition systems we consider are non-blocking.
Let us remark that this is the case for transitions systems $T_\tau(\Sigma)$ and $T_{\tau,\eta}(\Sigma)$ considered in this paper.

\begin{definition}
Let $\mathcal C$ be a controller for $T=(X,U,\Succ,Y,\Out)$ such that for all $x\in X$, $\mathcal C(x) \ne \emptyset$.
The {\it entry time of $T/{\mathcal C}$ from $x^0\in X$ for reachability specification $(Y_S,Y_T)$} 
is the smallest $N\in \N$ such that for all state trajectories of  $T/{\mathcal C}$, of length $N$ and starting from $x^0$,
$(x^0,u^0),(x^1,u^1),\dots,(x^{N-1},u^{N-1}),(x^N,u^N)$, there exists $K\in \{0,\dots,N\}$ such that
\begin{enumerate}
\item $\forall k \in \{0,\dots,K\},\; \Out(x^k) \in Y_S$;
\item $ \Out(x^K)\in Y_T$.
\end{enumerate}
The entry time is denoted by $J(T/{\mathcal C},Y_s,Y_t,x^0)$.
If such a $N\in \N$ does not exist, then we define $J(T/{\mathcal C},Y_S,Y_T,x^0)= + \infty$.
\end{definition}

It is clear  from the previous definition that for any initial state $x^0$ with finite entry time, the outputs of the controlled system $T/{\mathcal C}$ will remain in the safe set $Y_S$ until one output eventually reaches the target set $Y_T$ in a number of transitions bounded by $J(T/{\mathcal C},Y_S,Y_T,x^0)$.  Hence, for those states, the reachability specification is met.
It should be noted that for all $x^0\in X$,
 $J(T/{\mathcal C},Y_S,Y_T,x^0)=0$ if and only if $\Out(x^0) \in Y_T$
 and that for all $x^0\in X$ such that $\Out(x^0) \notin Y_S$,
 $J(T/{\mathcal C},Y_S,Y_T,x^0)=+\infty$.  Also for all $x\in X$, such that $0<J(T/{\mathcal C},Y_S,Y_T,x)<+\infty$, it is easy to show that
\begin{equation}
\label{eq:reach}
 J(T/{\mathcal C},Y_S,Y_T,x) = 1+ \max_{u\in {\mathcal C}(x), x'\in {\mathcal S}(x,u)}J(T/{\mathcal C},Y_S,Y_T,x').
 \end{equation}

We now consider the problem of synthesizing a reachability controller for $T_\tau(\Sigma)$ describing the sampled dynamics of the switched system $\Sigma$.
Let us consider a reachability specification given by compact sets $Y_S \subseteq \R^n$ and $Y_T\subseteq Y_S$.

\begin{thm}
\label{th:quantized2}
Let $\mathcal K_\varepsilon : [\R^n]_\eta \rightarrow 2^P$ be a controller for the symbolic model $T_{\tau,\eta}(\Sigma)$, let the map $\mathcal K: [\R^n]_\eta \rightarrow 2^P$ be given
for $q\in [\R^n]_\eta$ by\;\footnote{The function $\argmin$ is to be understood as a set-valued map: it returns the set of minimizers.} 
\begin{equation}
\label{eq:ref2}
\mathcal K(q) =  \mathcal K_\varepsilon \left(\argmin_{q'\in \overline{\mathcal R}_\varepsilon (q)} 
J(T_{\tau,\eta}(\Sigma)/{\mathcal K_\varepsilon},\textrm{Cont}_{\varepsilon}(Y_S),\textrm{Cont}_{\varepsilon}(Y_T),q')\right).
\end{equation}
Then, the map $\mathcal C:\R^n \rightarrow 2^P$ given by
$\mathcal C =\mathcal K \circ Q_\eta$ satisfies for all $x\in \R^n$:
\begin{equation}
\label{eq:ref4}
J(T_{\tau}(\Sigma)/{{\mathcal C}},Y_S,Y_T,x) \le {\tilde J}(Q_\eta(x))
\end{equation}
where $\tilde J:[\R^n]_\eta \rightarrow \N$ is the map given 
for $q\in [\R^n]_\eta$ by
$$
 {\tilde J}(q)=
\min_{q'\in \overline{\mathcal R}_\varepsilon (q)} J(T_{\tau,\eta}(\Sigma)/{\mathcal K_\varepsilon},\textrm{Cont}_{\varepsilon}(Y_S),\textrm{Cont}_{\varepsilon}(Y_T),q').
$$
\end{thm}

\begin{pf}
By Theorem~3 in~\cite{Girard2012}, we have that $\mathcal C : \R^n \rightarrow 2^P$ given by
\begin{equation}
\label{eq:1}
\mathcal C(x) =  \mathcal K_\varepsilon \left(\argmin_{q'\in {\mathcal R}_\varepsilon (x)} 
J(T_{\tau,\eta}(\Sigma)/{\mathcal K_\varepsilon},\textrm{Cont}_{\varepsilon}(Y_S),\textrm{Cont}_{\varepsilon}(Y_T),q')\right).
\end{equation}
satisfies
\begin{equation}
\label{eq:2}
J(T_{\tau}(\Sigma)/{{\mathcal C}},Y_S,Y_T,x) \le \min_{q'\in {\mathcal R}_\varepsilon (x)} J(T_{\tau,\eta}(\Sigma)/{\mathcal K_\varepsilon},\textrm{Cont}_{\varepsilon}(Y_S),\textrm{Cont}_{\varepsilon}(Y_T),q').
\end{equation}
Then, using the fact that $\mathcal R_\varepsilon(x)= \overline{\mathcal R}_\varepsilon(Q_\eta(x))$, 
equation (\ref{eq:1}) gives $\mathcal C=\mathcal K \circ Q_\eta$ and equation (\ref{eq:2}) gives (\ref{eq:ref4}).
$\blacksquare$
\end{pf}

Similarly to safety controllers, the synthesis of a reachability controller $\mathcal K_\varepsilon$ for the symbolic model $T_{\tau,\eta}(\Sigma)$ can be done by a simple fixed-point algorithm (e.g. using dynamic programming) which is guaranteed to terminate in a finite number of steps since $Y_S$ is compact.
It should be noted that we are only interested in the values of $\mathcal C(x)$ for $x\in Y_S$ since from $x\notin Y_S$ the reachability specification cannot be met.
Hence, it is only necessary to compute $\mathcal K$ on $Q_\eta(Y_S)$ which is finite since $Y_S$ is a compact subset of $\R^n$.
Therefore, the map $\mathcal K$ can  be pre-computed offline.
Thus, Theorem~\ref{th:quantized2} gives an effective way to compute a quantized reachability controller
for $T_\tau(\Sigma)$. Moreover, it is possible to give guarantees on the distance between the performances of the synthesized controller $\mathcal C$ and the time optimal  controller for the reachability specification $(Y_S,Y_T)$~\cite{Girard2012}.
The complexity of the synthesized controller in terms of execution time and memory consumption is similar to that of the safety controllers discussed in the previous section.


\begin{rmk}
We would like to highlight some relations between the control problems under consideration in this paper and some problems in viability theory~\cite{aubin}.
For the safety controller $\mathcal K$ defined in Theorem~\ref{th:quantized}, it can be shown that $\textrm{dom}(\mathcal K)$ is an under-approximation of the viability kernel of $\mathcal O_1^{-1}(Y_S)$ under the dynamics of $\Sigma$. As for the reachability controller $\mathcal K$ defined in Theorem~\ref{th:quantized2}, it can be shown that the set $\{x \in \R^n |  {\tilde J}(Q_\eta(x)) <+\infty \}$ is an under-approximation of the viable capture basin of $\mathcal O_1^{-1}(Y_T)$ in $\mathcal O_1^{-1}(Y_S)$
under the dynamics of $\Sigma$. 
\end{rmk} 
 
\section{Complexity Reduction}

We now consider the problem of representing the discrete maps $\mathcal K$ defined in Theorems~\ref{th:quantized}  and~\ref{th:quantized2} more efficiently in order to reduce the memory space needed for their storage.
To reduce the memory needed to store the control law, we will not encode the (set-valued) maps
$\mathcal K$ but  {\it determinizations} of $\mathcal K$.

\subsection{Determinization of safety controllers}

We first explain our approach for safety controllers. 
Let ${\mathcal K}$ be the map defined in Theorem~\ref{th:quantized} and let ${\mathcal C}=\mathcal K\circ Q_\eta$.

\begin{definition}
\label{def:det} A {\it determinization} of the set-valued map $\mathcal K$ is a univalued map $\mathcal K_d: Q_\eta(Y_S) \rightarrow P$ such that
$$
\forall q\in Q_\eta(Y_S) ,\; \mathcal K(q) \ne \emptyset \Rightarrow \mathcal K_d(q) \in  \mathcal K(q).
$$
\end{definition}

If $\mathcal K(q) = \emptyset$, we do not impose any constraint on the value of $\mathcal K_d(q)$.
This will allow us to reduce further the complexity of our control law. 

\begin{thm}\label{th:det} Let the controller $\mathcal C_d:\R^n \rightarrow 2^P$ for $T_\tau(\Sigma)$ be given for all $x \in \R^n$ by
$$
\mathcal C_d(x)=\left\{
\begin{array}{cl}
\{\mathcal K_d (Q_\eta(x))\} & \text{if } x \in Y_S \\
\emptyset & \text{otherwise.}
\end{array}
\right.
$$
Then, for all state trajectories $\{(x^i,u^i)|\; i=0,\dots,N\}$ of the controlled system $T_\tau(\Sigma)/{\mathcal C_d}$ such that $x^0\in \textrm{dom}(\mathcal C)$, we have $\mathcal O_1(x^i) \in Y_S$ for all $i=0,\dots,N$ and if $N$ is finite $x_N$ is a non-blocking state of $T_\tau(\Sigma)/{\mathcal C_d}$.
\end{thm}

\begin{pf} Since ${\mathcal C}$ is a safety controller we have $\textrm{dom}(\mathcal C) \subseteq Y_S = \textrm{dom}(\mathcal C_d)$. 
Let  $x\in \textrm{dom}(\mathcal C)$, then $x\in \textrm{dom}(\mathcal C_d)$ and therefore
$x$ is a non-blocking state of $T_\tau(\Sigma)/{\mathcal C_d}$. 
Let $p\in \mathcal C_d(x)$, since $\mathcal K(Q_\eta(x))={\mathcal C}(x)  \ne \emptyset$, Definition~\ref{def:det} implies that
$p=\mathcal K_d(Q_\eta(x)) \in \mathcal K(Q_\eta(x))=  \mathcal C(x)$. 
Since $\mathcal C$ is a safety controller, it follows that $x'=\mathcal S_1(x,p) \in \textrm{dom}(\mathcal C)$.
From the previous discussion, it follows by induction that for all $i=0,\dots,N$, $x^i \in  \textrm{dom}(\mathcal C)$. Moreover, if $N$ is finite 
$x_N$ is a non-blocking state of $T_\tau(\Sigma)/{\mathcal C_d}$. Finally,  since $\mathcal C$ is a safety controller, $x^i \in  \textrm{dom}(\mathcal C)$
gives $\mathcal O_1(x^i) \in Y_S$ for all $i=0,\dots,N$.
$\blacksquare$
\end{pf}
 
Let us remark that the controller $\mathcal C_d$ is generally not a safety controller for  $T_\tau(\Sigma)$ and specification $Y_S$
in the sense of Definition~\ref{def:safe} because there might be states in $ \textrm{dom}(\mathcal C_d)$ for which  the safety specification is not met. 
However, the previous result shows that for an initial state $x^0\in \textrm{dom}(\mathcal C)$, the controlled system
$T_\tau(\Sigma)/{\mathcal C_d}$ will never reach a blocking state and its outputs will remain forever in the safe set $Y_S$.

\subsection{Determinization of reachability controllers}

We now do a similar work for reachability controllers. Let ${\mathcal K}$ and ${\tilde J}$ be the maps defined in Theorem~\ref{th:quantized2} and let  ${\mathcal C}=\mathcal K\circ Q_\eta$.

\begin{definition}
\label{def:det2} A {\it determinization} of the set-valued map $\mathcal K$ is a univalued map $\mathcal K_d: Q_\eta(Y_S) \rightarrow P$ such that
$$
\forall q\in Q_\eta(Y_S\setminus Y_T) ,\;  {\tilde J}(q) < +\infty  \Rightarrow \mathcal K_d(q) \in  \mathcal K(q).
$$
\end{definition}

If ${\tilde J}(q) = +\infty $, or if $q\notin   Q_\eta(Y_S\setminus Y_T)$, we do not impose any constraint on the value of $\mathcal K_d(q)$.
This will allow us to reduce further the complexity of our control law. 

\begin{thm}\label{th:det2} Let the controller $\mathcal C_d:\R^n \rightarrow 2^P$ for $T_\tau(\Sigma)$ be given for all $x \in \R^n$ by
$$
\mathcal C_d(x)=\left\{
\begin{array}{cl}
\{\mathcal K_d (Q_\eta(x))\} & \text{if } x \in Y_S\setminus Y_T \\
P & \text{otherwise.}
\end{array}
\right.
$$
Then, for all $x\in \R^n$, 
\begin{equation}
\label{eq:ref14}
J(T_{\tau}(\Sigma)/{{\mathcal C_d}},Y_S,Y_T,x) \le  {\tilde J}(Q_\eta(x)).
\end{equation}
\end{thm}

\begin{pf}  If $x\notin Y_S$,  it follows that $J(T_{\tau}(\Sigma)/{{\mathcal C_d}},Y_S,Y_T,x)=+\infty$ and that $J(T_{\tau}(\Sigma)/{{\mathcal C}},Y_S,Y_T,x)=+\infty$. Then,
equation (\ref{eq:ref4}) gives $ {\tilde J}(Q_\eta(x))=+\infty$ and (\ref{eq:ref14}) holds. If $x\in Y_S$ and $ {\tilde J}(Q_\eta(x))=+\infty$ then (\ref{eq:ref14}) clearly holds as well. 
The only remaining case is $x\in Y_S$ and $ {\tilde J}(Q_\eta(x))<+\infty$. 
We now proceed by induction to show that
\begin{equation}
\label{eq:ref15}
J(T_{\tau}(\Sigma)/{{\mathcal C_d}},Y_S,Y_T,x) \le J(T_{\tau}(\Sigma)/{{\mathcal C}},Y_S,Y_T,x)
\end{equation}
which together with equation (\ref{eq:ref4}) gives (\ref{eq:ref14}). The induction is on the value of $J(T_{\tau}(\Sigma)/{{\mathcal C_d}},Y_S,Y_T,x)$.
Let $x$ be such that $J(T_{\tau}(\Sigma)/{{\mathcal C_d}},Y_S,Y_T,x)=0$, then $x\in Y_T$ and $J(T_{\tau}(\Sigma)/{{\mathcal C}},Y_S,Y_T,x)=0$ as well.
Let us assume that there exists $N\in \N$ such that for all $x$ such that $J(T_{\tau}(\Sigma)/{{\mathcal C_d}},Y_S,Y_T,x)\le N$, equation (\ref{eq:ref15}) holds.
We have shown that it is satisfied for $N=0$. Then, let $x$ such that $J(T_{\tau}(\Sigma)/{{\mathcal C_d}},Y_S,Y_T,x)= N+1$.
Then, we have $0<J(T_{\tau}(\Sigma)/{{\mathcal C_d}},Y_S,Y_T,x)<+\infty$ which implies that $x\in Y_S\setminus Y_T$.
Moreover, since $ {\tilde J}(Q_\eta(x))<+\infty$, we have
by Definition~\ref{def:det2} and by construction of $\mathcal C_d$, that ${\mathcal C}_d(x) \subseteq {\mathcal C}(x)$.
Let $p \in  {\mathcal C}_d(x)$ and $x'\in {\mathcal S}_1(x,p)$, then equation (\ref{eq:reach}) gives that  $J(T_{\tau}(\Sigma)/{{\mathcal C_d}},Y_S,Y_T,x')\le N$. 
Then, the induction assumption gives $J(T_{\tau}(\Sigma)/{{\mathcal C_d}},Y_S,Y_T,x')\le J(T_{\tau}(\Sigma)/{{\mathcal C}},Y_S,Y_T,x')$.
Then, equation (\ref{eq:reach}) yields
\begin{eqnarray*}
 J(T_{\tau}(\Sigma)/{\mathcal C_d},Y_S,Y_T,x) & = & 1+ \max_{p\in {\mathcal C}_d(x), x'\in {\mathcal S}(x,p)}J(T_{\tau}(\Sigma)/{\mathcal C_d},Y_S,Y_T,x') \\
 & \le &  1+ \max_{p\in {\mathcal C}_d(x), x'\in {\mathcal S}(x,p)}J(T_{\tau}(\Sigma)/{\mathcal C},Y_S,Y_T,x') \\
 & \le &  1+ \max_{p\in {\mathcal C}(x), x'\in {\mathcal S}(x,p)}J(T_{\tau}(\Sigma)/{\mathcal C},Y_S,Y_T,x') \\
 & \le & J(T_{\tau}(\Sigma)/{\mathcal C},Y_S,Y_T,x). 
\end{eqnarray*}
This completes the induction.
$\blacksquare$
\end{pf}

The previous result essentially states that using the controller ${\mathcal C}_d$,  the reachability specification will be met 
for all initial states $x^0\in Y_S$, such that ${\tilde J}(Q_\eta(x))<+\infty$. Moreover, equation (\ref{eq:ref15}) shows that  from those initial states, the entry time
using the controller  ${\mathcal C}_d$ cannot be larger than the entry time using the controller  ${\mathcal C}$.

\subsection{Efficient representation using algebraic decision diagrams}

We now consider the problem of choosing an appropriate determinization $\mathcal K_d$ of $\mathcal K$ and a representation 
which requires little memory for its storage. We explain our approach for safety controllers but it can be extended in a straightforward manner to handle reachability controllers as well.
A natural representation for $\mathcal K_d$ would be to use an array
which would require $O(\eta^{-n})$ memory space. We propose a more efficient representation inspired by algebraic decision diagrams (ADD's). 
The main idea is to use a tree structure which exploits redundant information to represent the map in a more compact way.
Also in our case, when $\mathcal K(q)$ is empty or when it has more than $2$ elements, we have some flexibility for the choice of
$\mathcal K_d(q)$ which can be used to reduce the size of the representation.

The proposed method for choosing $\mathcal K_d$ essentially works as follows: if there exists $p\in P$ such that for all $q\in Q_\eta(Y_S)$, $\mathcal K(q) = \emptyset$ or $p\in \mathcal K(q)$,
we can choose $\mathcal K_d$ to be the map with constant value $p$ on $Q_\eta(Y_S)$. The memory space needed to store $\mathcal K_d$ is then $O(1)$.
If such an input value does not exist, then we can split (typically using a hyperplane) the set
$Q_\eta(Y_S)$ into $2$ subsets of similar sizes. This process can then be repeated iteratively:  we try to find a suitable constant value on each of the subsets and
if this is not possible these sets can be split further.

\begin{figure}[!h]
\begin{center}
\includegraphics{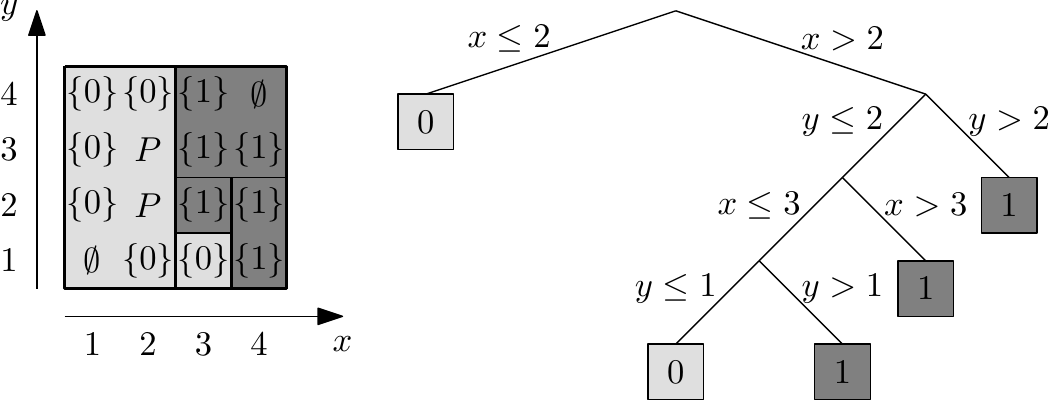}    
\caption{A set valued map $\mathcal K:\{1,2,3,4\}^2 \rightarrow 2^P$ where $P=\{0,1\}$ and a determinization given by colors (dark gray for $1$, light gray for $0$) and its representation using a tree structure.}  
\label{fig:graph}                                 
\end{center}                                 
\end{figure}

In Figure~\ref{fig:graph}, we show an example of representation using a tree structure of a determinization of a set-valued map $\mathcal K:\{1,2,3,4\}^2 \rightarrow 2^P$ where $P=\{0,1\}$. We cannot find a suitable constant value on the whole set $\{1,2,3,4\}^2$. Thus, it is split into two subsets
$\{1,2\}\times \{1,2,3,4\}$ and $\{3,4\}\times \{1,2,3,4\}$. For $q\in \{1,2\}\times \{1,2,3,4\}$ we can choose $\mathcal K_d(q)=0$.
On $\{3,4\}\times \{1,2,3,4\}$, there is no suitable value. This set is split further into the subsets $\{3,4\}\times \{1,2\}$ and $\{3,4\}^2$.
For $q\in\{3,4\}^2$, we can choose $\mathcal K_d(q)=1$. On $\{3,4\}\times \{1,2\}$, there is no suitable value and this set has to be split futher...
By repeating this process, we obtain the determinization $\mathcal K_d$ represented by the tree structure in Figure~\ref{fig:graph}.

\begin{rmk}
For reachability controllers, the approach is essentially the same except that for all region in our partition there must be a mode $p\in P$ such 
that for all $q$ in the region $\tilde J(q)=+\infty$ or $q\notin Q_\eta(Y_S\setminus Y_T)$ or $p\in \mathcal K(q)$.
\end{rmk}

Using this representation for the determinization $\mathcal K_d$, the online execution time of the controller $\mathcal C_d$ is given by the 
longest path in the tree which is in $O(-n\log(\eta))$. This is a little bit more than the execution time of controller $\mathcal C$. The memory space needed to store
the control law is given by the number of nodes in the tree which is $O(\eta^{-n})$, in the worst case. However, in practice, we can expect much less
as an example will show in the next section.

Finally, we would like to mention that the use of binary decision diagrams (a special class of ADD's) for representing control laws synthesized through symbolic models has already been considered
in~\cite{Pessoa}. However, as far as we know, the idea of determinizing  controllers 
in such a way that their determinization reduces the memory needed for its storage is new.

\section{Example}

For illustration purpose, we consider a simple thermal model of a two-room building (see e.g~\cite{building}):
$$
\hspace{-0.4cm}
\left\{
\begin{array}{lll}
\dot T_1 & = & \alpha_{21} (T_2-T_1) + \alpha_{e1} (T_e-T_1) + \alpha_{f} (T_f-T_1) p \\
\dot T_2 & = & \alpha_{12} (T_1-T_2) + \alpha_{e2} (T_e-T_2)
\end{array}
\right.
$$
where $T_1$ and $T_2$ denote the temperature in each room, $T_e=10$ is the external temperature and $T_f$ stands for the temperature of a heating device which can switched on ($p=1$) or off ($p=0$).  
The system parameters are chosen as follows $\alpha_{21}=\alpha_{12}=5\times 10^{-2}$,
$\alpha_{e1}= 5\times 10^{-3}$, $\alpha_{e2}= 3.3 \times 10^{-3}$ and $\alpha_f = 8.3 \times 10^{-3}$.
Let $T=(T_1,T_2)^\top$, then the system can be written as a switched affine system of the form
$$
\Sigma:\; \dot {\mathbf T}(t) = A_{{\mathbf p}(t)}{\mathbf T}(t)+b_{{\mathbf p}(t)},\; {\mathbf p}(t)\in P=\{0,1\}.
$$
It is easily to verify that the function $\mathcal V: \R^2\times \R^2 \rightarrow \R^+_0$ given by $\mathcal V(T,T')=\|T-T'\|^2$ is a $\delta$-GUAS Laypunov function
for $\Sigma$ with $\underline \alpha(r)=\overline \alpha(r)=r^2$ and $\kappa=0.0084$. Moreover, equation~(\ref{eq:assum}) holds with $\gamma(r)=r^2$.

We first consider the problem of keeping the temperature in the rooms between $20$ and $22$ degrees Celsius. This is a safety property specified by the safe set $Y_S=[20,22]^2$. We want to use a periodic controller with a period of $\tau=5$ time units. For the synthesis of the controller, we shall use an
approximately bisimilar symbolic abstraction of $T_{\tau}(\Sigma)$ of precision $\varepsilon=0.25$. According to equation (\ref{eq:simcond}), we can choose a state-space sampling parameter $\eta=0.0014$ for the computation of the symbolic abstraction $T_{\tau,\eta}(\Sigma)$.

\begin{figure}[!t]
\begin{center}
\hspace{-0.2cm}
\includegraphics[width=0.52\columnwidth]{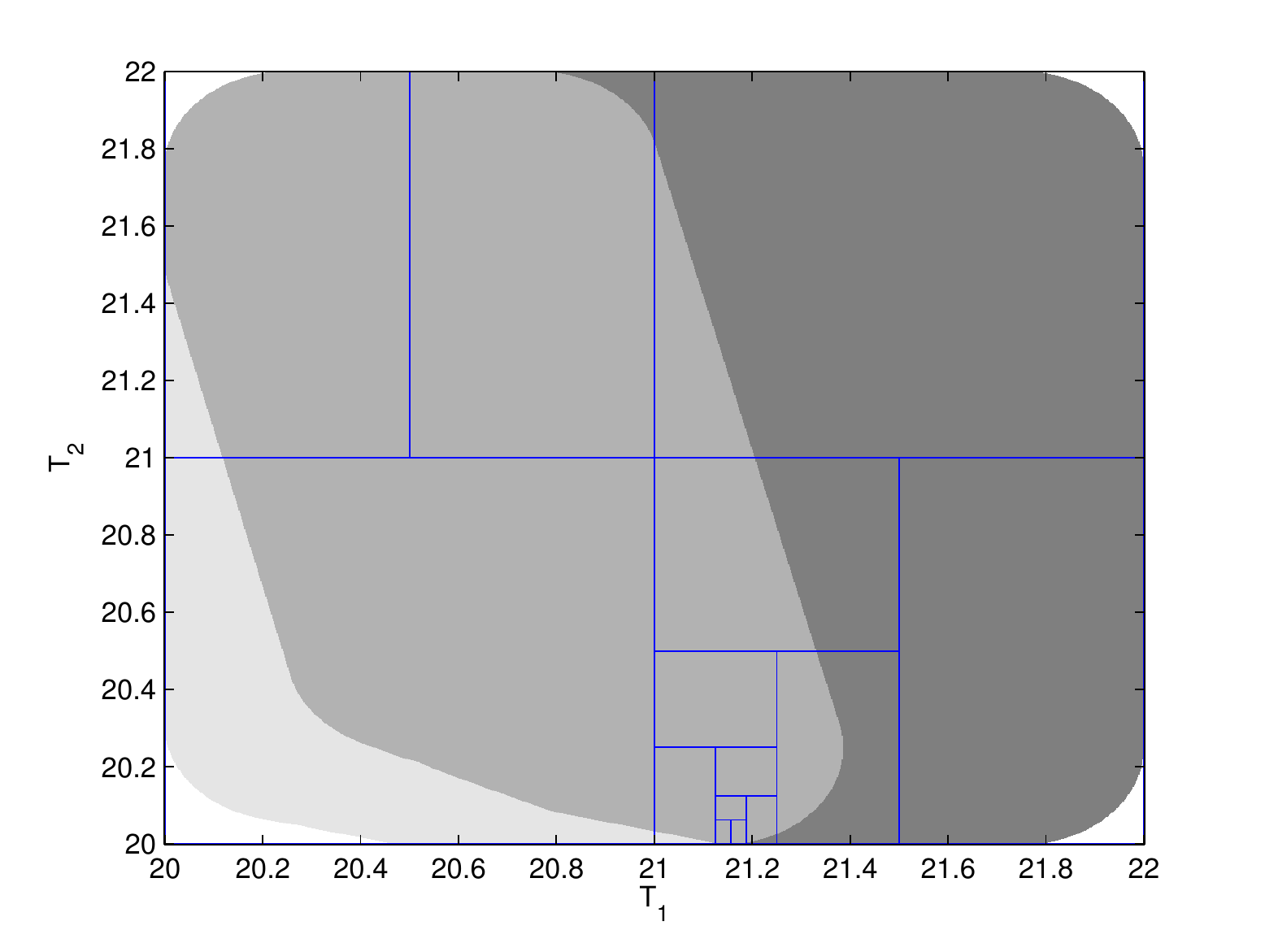}  
\hspace{-0.7cm}
\includegraphics[width=0.52\columnwidth]{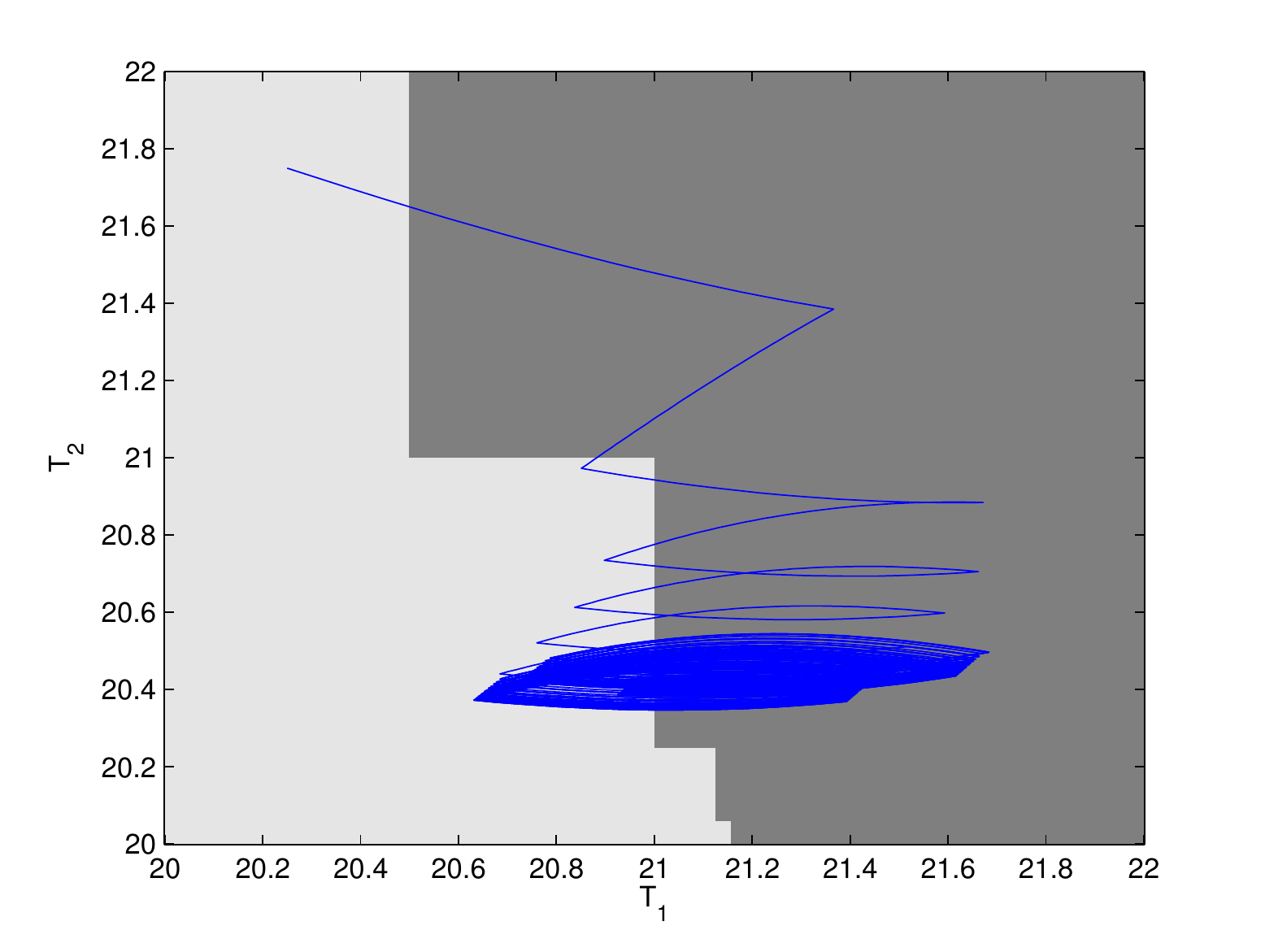} 
\caption{Left: Set-valued map $\mathcal K: Q_\eta( Y_S) \rightarrow 2^P$ (white: $\emptyset$, light gray: $\{1\}$, medium gray: $P$, dark gray: $\{0\}$).
The number of elements in $Q_\eta( Y_S) $ is about $1$ million. In blue, we represented the partition used for the representation of $\mathcal K_d$, a determinization of $\mathcal K$; the resulting tree structure has only $27$ nodes. Right:
Determinization $\mathcal K_d$ of the map $\mathcal K$ shown on the left (light gray: $1$, dark gray: $0$). In blue, a trajectory of the switched system controlled using the controller $\mathcal C_d=\mathcal K_d \circ Q_\eta$.
}  
\label{figC}                                 
\end{center}                                 
\end{figure}

We computed a safety controller $\mathcal K_\varepsilon$ for the symbolic abstraction $T_{\tau,\eta}(\Sigma)$ and the specification
$\textrm{Cont}_\varepsilon (Y_S)=[20.25,21.75]^2$. Then, we computed the map $\mathcal K$ given by equation~(\ref{eq:ref}),
which is shown in the left part of Figure~\ref{figC}. Then, according to Theorem~\ref{th:quantized},  
the controller $\mathcal C=\mathcal K \circ Q_\eta$ is a safety controller for $T_\tau(\Sigma)$ and specification $Y_S$.
For a practical implementation of the controller, the storage of the map $\mathcal K$ represented by an array would require about $1$
million memory units (this is the number of elements in $Q_\eta(Y_S)$).
%
%
%
%
We computed a determinization $\mathcal K_d$ of $\mathcal K$ following the approach described in the previous section.
In Figure~\ref{figC}, we show the partition used for the representation of $\mathcal K_d$, it is to be noted that in each region all values of
$\mathcal K$ are either $\emptyset$, $\{0\}$, $P$ (which corresponds to value $0$ for $\mathcal K_d$) or $\emptyset$, $\{1\}$, $P$ 
(which corresponds to value $1$ for $\mathcal K_d$).
The map $\mathcal K_d$ is
represented in the right part of Figure~\ref{figC} where we have also represented a trajectory of the switched system controlled using the controller $\mathcal C_d$. For a practical implementation of the controller, the storage of the map $\mathcal K_d$ represented by a tree structure only requires $27$ memory units (this is the number of nodes in the tree).
We can see with this example that a lot of memory can be saved using an efficient representation and by determinizing the controllers
in such a way that their determinization can be represented in a more compactly. Guarantees of safety for these controllers are still available 
by Theorem~\ref{th:det} which gives insurance of ``correctness by design''.

We now consider the problem of setting the temperature in the rooms between $20$ and $22$ degrees Celsius while keeping it between $17.5$ and $22.5$ along the way.
This a reachability specification with $Y_S=[17.5,22.5]^2$ and $Y_T=[20,22]^2$.  For the synthesis of the controller, we shall use an
approximately bisimilar symbolic abstraction of $T_{\tau}(\Sigma)$ of precision $\varepsilon=0.5$. According to equation (\ref{eq:simcond}), we can choose a state-space sampling parameter $\eta=0.0035$ for the computation of the symbolic abstraction $T_{\tau,\eta}(\Sigma)$.

\begin{figure}[!t]
\begin{center}
\hspace{-0.2cm}
\includegraphics[width=0.52\columnwidth]{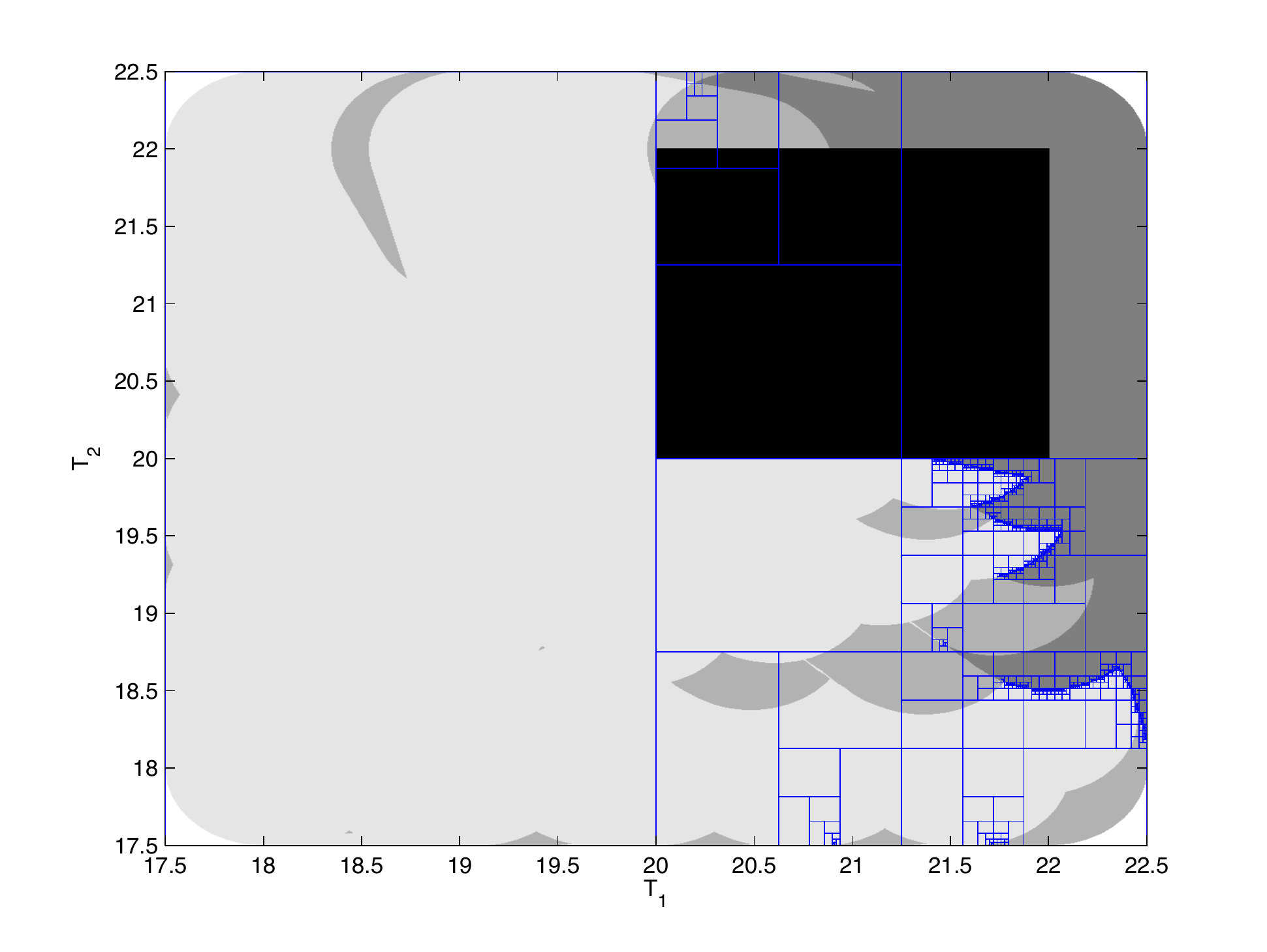}  
\hspace{-0.7cm}
\includegraphics[width=0.52\columnwidth]{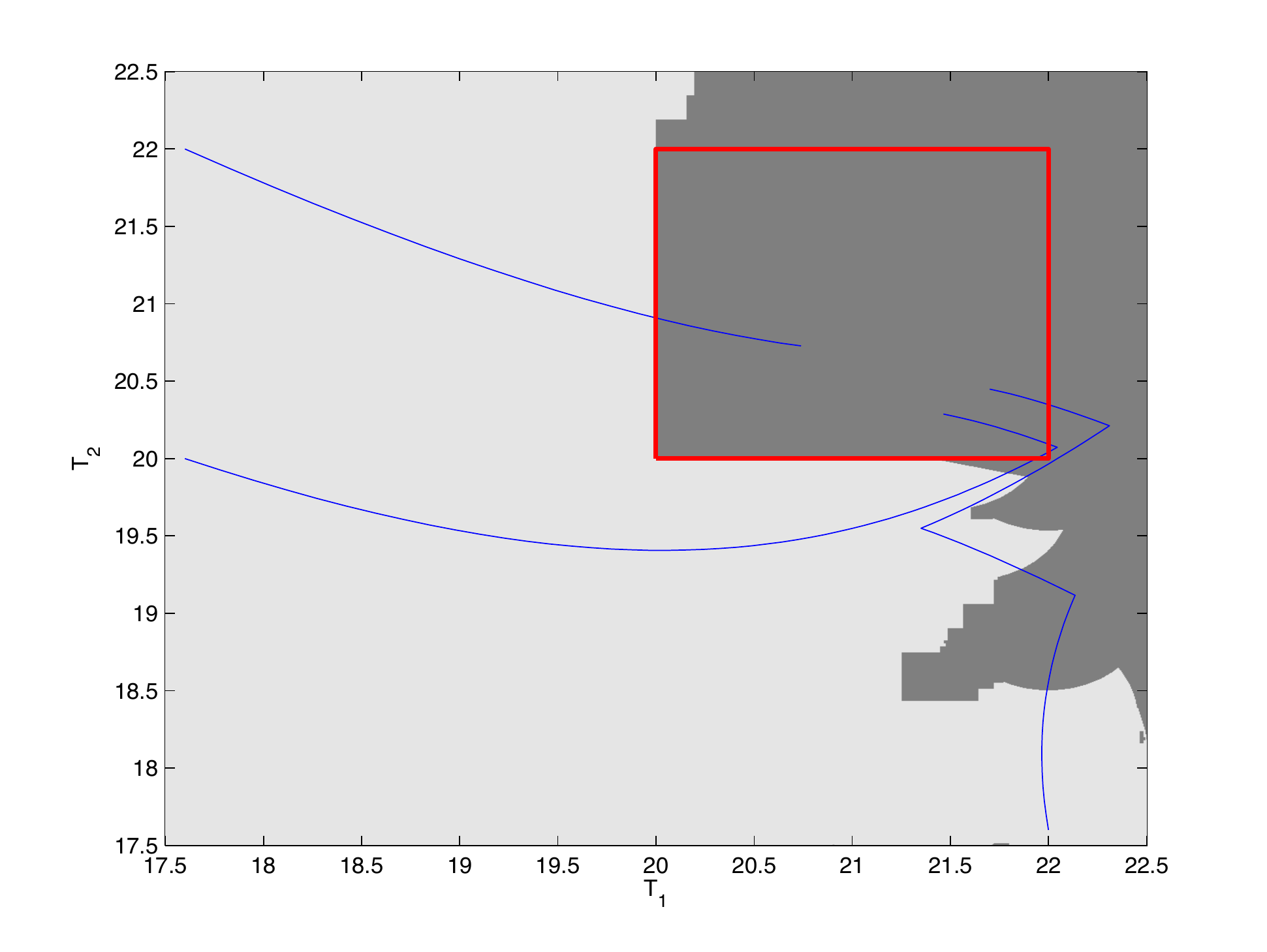} 
\caption{Left: Set-valued map $\mathcal K: Q_\eta( Y_S) \rightarrow 2^P$ (light gray: $\{1\}$, medium gray: $P$, dark gray: $\{0\}$, white: $\tilde J(q)=+\infty$, black: $q\notin Q_\eta(Y_S\setminus Y_T)$).
The number of elements in $Q_\eta( Y_S) $ is about $1$ million. In blue, we represented the partition used for the representation of $\mathcal K_d$, a determinization of $\mathcal K$; the resulting tree structure has  $2249$ nodes. Right:
Determinization $\mathcal K_d$ of the map $\mathcal K$ shown on the left (light gray: $1$, dark gray: $0$). In blue, a trajectory of the switched system controlled using the controller $\mathcal C_d=\mathcal K_d \circ Q_\eta$.
}  
\label{figD}                                 
\end{center}                                 
\end{figure}

We computed a reachability controller $\mathcal K_\varepsilon$ for the symbolic abstraction $T_{\tau,\eta}(\Sigma)$ and the specification
$\textrm{Cont}_\varepsilon (Y_S)=[18,22]^2$, $\textrm{Cont}_\varepsilon (Y_T)=[20.5,21.5]^2$. Then, we computed the map $\mathcal K$ given by equation~(\ref{eq:ref2}),
which is shown in the left part of Figure~\ref{figD}. 
For a practical implementation of the controller, the storage of the map $\mathcal K$ represented by an array would require about $1$
million memory units.

We computed a determinization $\mathcal K_d$ of $\mathcal K$ following the approach described in the previous section.
In Figure~\ref{figD}, we show the partition used for the representation of $\mathcal K_d$.
The map $\mathcal K_d$ is
represented in the right part of Figure~\ref{figC} where we have also represented a trajectory of the switched system controlled using the controller $\mathcal C_d$. For a practical implementation of the controller, the storage of the map $\mathcal K_d$ represented by a tree structure only requires $2249$ memory units (this is the number of nodes in the tree).
Though the compression is not as spectacular as in the previous example $2249$ is still much less than $1$ million. Morover, Theorem~\ref{th:det2} gives insurance of ``correctness by design''. 

\section{Conclusion}

In this paper, we have addressed the problem of synthesizing low-complexity quantized controllers for switched systems for safety and reachability specifications. 
By following a rigorous approach based on the use of symbolic models we obtain controllers that are correct by design.
Determinization of the safety controllers together with an adequate data structure can reduce drastically the memory needed to store
the control law and can lead to quantized controllers that can be efficiently implemented in practice.

In future work, we should address the problem of synthesizing low-complexity controllers using other types of symbolic models such as multi-scale symbolic models introduced in~\cite{camara2011}.

\bibliographystyle{elsarticle-num}        
\bibliography{symcon}      

\end{document}